\documentclass[12pt]{article}
\usepackage{amsmath,amsfonts,epsfig,mathrsfs,todonotes,yfonts}
\usepackage{xcolor}
\usepackage{cite}
\usepackage{stmaryrd}
\usepackage{mdframed}
\usepackage{scalerel}
\usepackage[top=2.5cm, bottom=2.5cm, left=2.75cm, right=2.75cm]{geometry} 
\usepackage{dsfont}
\numberwithin{equation}{section}
\usepackage{hyperref}
 \hypersetup{
     colorlinks=true,
     linkcolor=black,
     filecolor=blue,
     citecolor=blue,      
     urlcolor=cyan,
     }
\usepackage[most]{tcolorbox}

\newcommand{\re}[1] {(\ref{#1})}

\newcommand{\pa}{\partial} % curly d
\newcommand{\al}{\alpha}

\newcommand{\Om}{\Omega}

\newcommand{\ber}{\begin{eqnarray}}
\newcommand{\eer}[1]{\label{#1}\end{eqnarray}}
\newcommand{\eero}{\end{eqnarray}}

\newcommand{\balg}{\begin{align}}
\newcommand{\ealg}{\end{align}}

\newcommand{\beq}{\begin{equation}}
\newcommand{\eeq}{\end{equation}}
\newcommand{\bea}{\begin{eqnarray}}
\newcommand{\eea}{\end{eqnarray}}

\newcommand{\rd}[1]{{\color{red}{#1}}}

\newcommand{\nn}{\nonumber}
\newcommand{\na}{\nabla}

\newcommand{\half}{{\textstyle{\frac12}}}

\def\HollowBox #1#2{{\dimen0=#1 \advance\dimen0 by -#2
       \dimen1=#1 \advance\dimen1 by #2
        \vrule height #1 depth #2 width #2
        \vrule height 0pt depth #2 width #1
        \llap{\vrule height #1 depth -\dimen0 width \dimen1} 
       \hskip -#2
       \vrule height #1 depth #2 width #2}}

\usepackage{fancyhdr}   
\newcommand{\auth}{\large 
{\large Ulf Lindstr\"om ${}^{a,b}$\footnote{email: ulf.lindstrom@physics.uu.se, Leverhulme Visiting Professor at Imperial College}}
and {\"O}zg{\"u}r Sar{\i}o\u{g}lu ${}^a$\footnote{email: sarioglu@metu.edu.tr}}
\begin{document}
%\pagestyle{fancy}
%\fancyhf{}
%\rhead{DRAFT}
\begin{flushright}
%{\small AEI-2009-087}\\
{\small UUITP-27/22}\\
%{\small Imperial-TP-UL-02}\\
\vskip 1.5 cm
\end{flushright}

\begin{center}
{\Large{\bf Geometry, conformal Killing-Yano tensors and conserved ``currents''}}
\vspace{.75cm}

\auth
\end{center}
\vspace{.5cm}
%\vspace{.5cm}
\centerline{${}^a${\it \small Department of Physics, Faculty of Arts and Sciences,}}
\centerline{{\it \small Middle East Technical University, 06800, Ankara, Turkey}}
\vspace{.5cm}
\centerline{${}^b${\it \small Department of Physics and Astronomy, Theoretical Physics, Uppsala University}}
\centerline{{\it \small SE-751 20 Uppsala, Sweden}}
\centerline{and}
\centerline{{\it \small Theoretical Physics, Imperial College London,}}
\centerline{{\it \small Prince Consort Road, London SW7 2AZ, UK}}

\vspace{1cm}

%\today

\centerline{{\bf Abstract}}
\bigskip

\noindent
{In this paper we discuss the construction of conserved tensors (currents) involving conformal Killing-Yano tensors (CKYTs), generalising the corresponding constructions for Killing-Yano tensors
(KYTs). As a useful preparation for this, but also of intrinsic interest, we derive identities relating 
CKYTs and geometric quantities. The behaviour of CKYTs under conformal transformations is 
also given, correcting formulae in the literature. We then use the identities derived to construct covariantly conserved ``currents''. We find several new CKYT currents and also include a known 
one by Penrose which shows that ``trivial'' currents are  also useful. We further find that  
rank-$n$ currents based on rank-$n$ CKYTs $k$ must have a simple form in terms of $dk$. 
By construction, these currents are covariant under a general conformal rescaling of the 
metric. How currents lead to conserved charges is then illustrated using the Kerr-Newman 
and the C-metric in four dimensions. Separately, we  study a rank-1 current, construct its charge 
and discuss its relation to the recently constructed Cotton current for the Kerr-Newman black 
hole.}
\vskip .5cm
 
\vspace{0.5cm}
\small

\renewcommand{\thefootnote}{\arabic{footnote}}
\setcounter{footnote}{0}

\pagebreak
\tableofcontents
\setcounter{page}{2}

\section{Introduction}\label{intro}
In recent papers \cite{Lindstrom:2021dpm,Lindstrom:2021qrk,Lindstrom:2022iec}, we have
investigated Killing tensors (KTs) and Killing-Yano tensors (KYTs), identified new geometric 
identities constraining the geometries that carry such tensors and shown the existence of 
covariantly conserved antisymmetric tensors (that we call ``currents'') as well as worked out 
their related charges and, separately, asymptotic charges. 
%Our inspiration originally came from analyzing the 
%current construction in \cite{Kastor:2004jk}. 
So far, we have only considered conformal Killing-Yano tensors (CKYTs) in connection with 
the Cotton currents in \cite{Lindstrom:2021dpm}, but a natural question to ask is which of the 
KYT currents can be generalised to include CKYTs. In the present paper, we first derive 
useful identities and relations for CKYTs, including the transformation properties of a CKYT 
under a rescaling of the metric, and then discuss a number of currents based on CKYTs. 
The existence of such currents and charges can be quite rewarding; e.g. one can then look 
for applications that lead to finding additional supersymmetries or construct asymptotic 
charges as already done for KYTs.

For the special case of rank-$n$ currents based on rank-$n$ CKYTs we find the unexpected 
result that the divergences of all currents can be expressed in a trivial form reminiscent of a 
topological current. In the language of differential forms, they only involve $dk$ of the $n$-form 
CKYT $k$. However this is by no means  obvious. We reach this conclusion from a general ansatz 
making use of a number of identities derived in the preparatory section.

The outline of the paper is as follows: Section \!\ref{rankn} contains a derivation of identities 
of general interest, in particular \re{LSn}, \re{nLSn} and \re{BoxLSn}, some of which are needed 
for the derivation of the rank-$n$ conserved currents in section \!\ref{ConCurn}. In section 
\!\ref{CKYTcur} we present our CKYT currents relegating the discussion of the rank-$n$ 
problems to section \!\ref{ConCurn} with the simpler $n=1$ and $n=2$ cases in 
appendix \!\ref{specase}. Section \!\ref{charges} consists of two subsections in which we 
illustrate how to find charges from two of the currents discussed; the nontrivial ``Einstein 
current" utilizing the Kerr-Newman metric and the ``trivial current" deploying the C-metric. 
Our conclusions and a discussion are contained in section \!\ref{disc}. In appendix \!\ref{KYCcur}, 
we review the Cotton current and prove the important property of covariance under conformal
transformations for it as well. 

Here we try to focus our attention on the discussion of currents, but a list of publications 
containing useful general background information about (C)KYTs and their applications can 
be found in \cite{Collect}. Finally it should be noted that throughout, the geometric setting is 
always curved (pseudo-)Riemannian geometry in $D$ dimensions with the Levi-Civita connection.

\section{Rank-\texorpdfstring{$n$}{n} identities and transformation properties}\label{rankn}
In this section we define a CKYT and derive identities involving CKYTs and geometric tensors. 
These identities are nontrivial and allow us to identify possible constituent terms of 
conserved currents and to arrive at a general ansatz for such a current. This ansatz applied to  
a rank-$n$ current involving a rank-$n$ CKYT is shown in section \!\ref{ConCurn} to reduce to 
a trivial current, in the sense mentioned in the introduction. We also examine the behaviour of
CKYTs under conformal transformations and correct the literature on this.

\subsection{Identities}\label{IDs}
A CKYT $k$ generalises a KYT in the way a conformal Killing vector generalises a 
Killing vector. A rank-$n$ CKYT $k$ can be defined \cite{Kashiwada} as an $n$-form that 
satisfies
\bea
\na_{a_{1}} k_{a_{2} \dots a_{n+1}} & = & \na_{[a_{1}} k_{a_{2} \dots a_{n+1}]} 
+ n \, g_{a_{1}[a_{2}} K_{a_{3} \dots a_{n+1}]} \,,  \label{idna} \\
K_{a_{1} \dots a_{n-1}} & := & \frac{1}{(D-n+1)} \, \na_{c} k^{c}{}_{a_{1} \dots a_{n-1}} \,. \label{idnb} 
\eea
It follows that
\ber
 \na_{a_{1}} K^{a_{1} \dots a_{n-1}} = 0 \,. 
\eer{idnn} 
The definition reduces to that of a rank-$n$ KYT $f$ by setting \( K^{a_{1} \dots a_{n-1}} = 0 \). 
An important identity for KYTs \cite{Kastor:2004jk} may be generalized to CKYTs as 
follows\footnote{It has been pointed out that \re{LSn} is derivable from (1.4) of \cite{Kashiwada}. 
We find \re{LSn} more accessible.} 
\begin{tcolorbox}[ams align]
\na_{a} \na_{b} k_{c_{1} \dots c_{n}} & = 
(-1)^{n+1} \frac{(n+1)}{2} \, R^{d}{}_{a[bc_{1}} \, k_{c_{2} \dots c_{n}] d} 
- (n+1) g_{a[b} \na_{c_{1}} K_{c_{2} \dots c_{n}]} \nn \\
& \quad 
+ n \na_{a} \left( g_{b[c_{1}} K_{c_{2} \dots c_{n}]} \right) \,. \label{LSn}
\end{tcolorbox}
Using \re{idna} and \re{LSn} in the commutator \( [\na_a, \na_b] \na_c k_{c_{1} \dots c_{n}} \), 
and contracting the index pairs $(a,c_{n})$ and $(b,c)$ on both sides of the resultant expression, 
we get
\begin{eqnarray}
&& (n-1) \left( \na^{b} R^{a}{}_{[c_{1}} \right) k_{c_{2} \dots c_{n-1}]ab}
+ \frac{1}{2} (\na^{a} R) k_{a c_{1} \dots c_{n-1}} 
+ R K_{c_{1} \dots c_{n-1}} 
+ (D-n) \, \Box K_{c_{1} \dots c_{n-1}}
\nn \\
&&- \frac{1}{3}(n-1)(n-2)(D-n-1) \, K^{ab}{}_{[c_{1} \dots c_{n-3}} 
\left( R_{c_{n-2} |a| c_{n-1}] b} + R_{c_{n-2} c_{n-1}] ab} \right) 
\nn \\
&& + (n-1) (D-n-2) \, K_{[c_{1} \dots c_{n-2}}{}^{a} R_{c_{n-1}]a} 
- (n-1) (D-n) \, \na_{a} \na_{[c_{1}} K^{a}{}_{c_{2} \dots c_{n-1}]} =  0 \,. 
\label{nLSnpre}
\end{eqnarray}
Since 
\ber 
 2 K^{ab}{}_{[c_{1} \dots c_{n-3}} R_{c_{n-2} |a| c_{n-1}] b} = K^{ab}{}_{[c_{1} \dots c_{n-3}}
R_{c_{n-2} c_{n-1}] ab} \,,
\eer{Rid}
we have
\ber 
&&- \frac{1}{3} \, K^{ab}{}_{[c_{1} \dots c_{n-3}} 
\left( R_{c_{n-2} |a| c_{n-1}] b} + R_{c_{n-2} c_{n-1}] ab} \right) 
= - \frac{1}{2} \, K^{ab}{}_{[c_{1} \dots c_{n-3}} R_{c_{n-2} c_{n-1}] ab} \,.
\eer{KRid}
Moreover
\ber
&& \na_{a} \na_{[c_{1}} K^{a}{}_{c_{2} \dots c_{n-1}]} = 
 [ \na_{a}, \na_{[c_{1}} ] K^{a}{}_{c_{2} \dots c_{n-1}]}  = R_{a[c_{1}} K^{a}{}_{c_{2} \dots c_{n-1}]} 
 - (n-2) R_{b [ c_{1} c_{2} |a|} K^{ab}{}_{c_{3} \dots c_{n-1}]} \nn \\[1mm]
&& \implies \na_{a} \na_{[c_{1}} K^{a}{}_{c_{2} \dots c_{n-1}]} = 
 K_{[c_{1} \dots c_{n-2}}{}^{a} R_{c_{n-1}]a} 
 - (n-2) R_{b [ c_{n-2} c_{n-1} |a|} K^{ab}{}_{c_{1} \dots c_{n-3}]} \,.
\eer{KRiem}
These can be used for simplifying the second and third lines of \re{nLSnpre} to arrive at
\begin{tcolorbox}[ams align]
&& (n-1) \left( \na^{b} R^{a}{}_{[c_{1}} \right) k_{c_{2} \dots c_{n-1}]ab}
+ \frac{1}{2} (\na^{a} R) k_{a c_{1} \dots c_{n-1}} 
+ (D-n) \, \Box K_{c_{1} \dots c_{n-1}} + R K_{c_{1} \dots c_{n-1}}  \nn \\[1mm]
&& -2 (n-1) \, K_{[c_{1} \dots c_{n-2}}{}^{a} R_{c_{n-1}]a} 
+ \frac{1}{2} (n-1)(n-2) \, K^{ab}{}_{[c_{1} \dots c_{n-3}} R_{c_{n-2} c_{n-1}] ab} = 0 \,. 
 \label{nLSn}
\end{tcolorbox}
On the other hand, by contracting the $(a,b)$ indices in \re{LSn}, we also have
\begin{tcolorbox}[ams align]
\Box k^{c_{1} \dots c_{n}} = \frac{(n-1)}{2} k^{ab [c_{1} \dots c_{n-2}} R^{c_{n-1} c_{n}]}{}_{ab}
- R^{a [ c_{1}} k_{a}{}^{c_{2} \dots c_{n}]} + (2n-D) \na^{[c_{1}} K^{c_{2} \dots c_{n}]} \,.
\label{BoxLSn}
\end{tcolorbox}
These may be combined to define the following current 
\begin{tcolorbox}[ams align]
F^{c_{1} \dots c_{n}} = \Box k^{c_{1} \dots c_{n}} - n \na^{[c_{1}} K^{c_{2} \dots c_{n}]} 
= \na_{a} \left( \na^{[a} k^{c_{1} \dots c_{n}]} \right) \,.
\label{theCurn}
\end{tcolorbox}\noindent
It is easy to show that this current is indeed covariantly conserved, i.e.
\ber
\na_{c_{1}} F^{c_{1} \dots c_{n}} = 0 \,.
\eer{triva}
This kind of current will be called ``trivial''. It is conserved because it is the covariant 
divergence of an antisymmetric tensor, not due to any other property of the CKYT. The 
combination \( \na_{a} T^{ac_{2} \dots c_{n}} \) will be divergence-free for any completely 
antisymmetric tensor $T$. 

\subsection{CKYTs under conformal transformations}\label{Weyl}
In this subsection we give the transformation properties of CKYTs under conformal 
transformations and show that the covariant-conservation of the trivial current \re{theCurn} 
is left-invariant under such transformations.
 
Under conformal transformations \( \tilde{g}_{ab} = \Om^{2} g_{ab} \), a generic rank-$n$ 
CKYT $k$  \re{idna} and its rank-$(n-1)$ companion $K$ \re{idnb} transform as
\ber
\tilde{k}^{a_{1} \dots a_{n}} = \Om^{1-n} k^{a_{1} \dots a_{n}} \quad \mbox{and} \quad
\tilde{K}_{a_{1} \dots a_{n-1}} = \Om^{n-1} K_{a_{1} \dots a_{n-1}} 
+ \Om^{n-2} (\na_{c} \Om) k^{c}{}_{a_{1} \dots a_{n-1}} \,.
\eer{WeylkK}
Note that for all $n\neq2$, the second part in the transformations of $\tilde{K}$ differs 
significantly from the formulae given in appendix A of \cite{Chervonyi:2015ima}.

With these, we find
\ber
\tilde{\na}^{[c} \tilde{k}^{a_{1} \dots a_{n}]} =
\Om^{-(n+1)} \na^{[c} k^{a_{1} \dots a_{n}]} 
+ (n+1) \Om^{-(n+2)} (\na^{[c} \Om) k^{a_{1} \dots a_{n}]} 
\eer{Wdelk}
leading to the transformation rule
\bea
& & \tilde{F}^{a_{1} \dots a_{n}} = \Om^{-(n+1)} F^{a_{1} \dots a_{n}}
+(D-n-1) \Om^{-(n+2)} (\na_{c} \Om) \na^{[c} k^{a_{1} \dots a_{n}]} \label{WFcur} \\
&& \qquad \qquad
+(n+1) \Om^{-(n+2)} \Big( \na_{c} \big( (\na^{[c} \Om) k^{a_{1} \dots a_{n}]} \big)
+(D-n-2)  \Om^{-1}  (\na_{c} \Om) (\na^{[c} \Om) k^{a_{1} \dots a_{n}]} \Big) \nn 
\eea
for a generic rank-$n$ current $F$. Since for any skew-symmetric rank-$(n+1)$ tensor $\tilde{T}$,
one has
\ber
[ \tilde{\na}_{a_1}, \tilde{\na}_{c}] \tilde{T}^{c a_{1} \dots a_{n}} = 
2 \tilde{\na}_{a_1} \tilde{\na}_{c} \tilde{T}^{c a_{1} \dots a_{n}} = 0 \,,
\eer{deldelT}
which can be shown by the expansion of the commutator, the symmetries of the curvature tensors 
and the first Bianchi identity, it follows straightforwardly that 
\( \tilde{\na}_{a_{1}} \tilde{F}^{a_{1} \dots a_{n}} = 0 \) by the identifications
\( \tilde{T}^{c a_{1} \dots a_{n}} = \tilde{\na}^{[c} \tilde{k}^{a_{1} \dots a_{n}]} \) and
\( \tilde{F}^{a_{1} \dots a_{n}} = \tilde{\na}_{c} \tilde{T}^{c a_{1} \dots a_{n}} \).
So the covariant-conservation of the trivial current $F$ is left invariant under 
conformal transformations.

\section{CKYT currents}\label{CKYTcur}
In this section we present cases when CKYTs give rise to conserved currents. The conserved 
currents are characterised by their rank, the rank of the CKYT involved and the number of 
derivatives. One of the currents presented is related to the Kastor-Traschen current (KT-current) 
\cite{Kastor:2004jk} to which we return below in the special case of a rank-2 CKYT and 
rank-2 current.

From the definitions \re{idnn} we see that a conserved current involving a rank-2 CKYT $k$ 
in $D=4$ dimensions is \cite{Penrose:1982wp}
\ber
j^{a} := \epsilon^{abcd} \na_{b} k_{cd} \,.
\eer{C1}
Here \( \na_{a} j^a = 0 \) follows easily from \re{idna}. In this case the current is rank-1.
In fact, since the Hodge dual $\star{k}$ of a CKYT $k$ is again a CKYT \cite{Cariglia:2003kf} 
\( \hat{k} := \star{k} \), one has
\ber
j^{a} = \na_{b} \hat{k}^{ab} \,,
\eer{Pen}
so that this is a trivial current as described at the end of subsection \!\ref{IDs}. It is used 
in \cite{Penrose:1982wp} as the starting point for a discussion of KTs vs KYTs, and quasilocal 
charges. 

In \cite{Lindstrom:2021dpm}, we showed that there are also other rank-1 
currents constructed out of the Cotton tensor $C$ and a rank-2 KYT $f$ or a rank-2 CKYT $k$, 
the Killing-Yano Cotton currents
\ber
J^a := C^{abc} f_{bc} \,, \qquad \tilde{J}^a := C^{abc} k_{bc} \,.
\eer{C2}
These are covariantly conserved in arbitrary $D \geq 3$ dimensions. In showing that the 
Cotton currents are divergence free, we use the fact that the Cotton tensor is traceless on 
all index pairs and satisfies 
\ber
C_{abc} = C_{a[bc]} \,, \quad 
C_{[abc]} = 0 \,, \quad
\na^a C_{abc} = 0 \,.
\eer{Ccond}
The conservation of the currents follow from an interplay between the geometry and 
(C)KYT properties. It is therefore useful to explore such relations in some detail, which 
we have done in subsection \!\ref{IDs} and shall elaborate further in section \!\ref{ConCurn}.
 
The Hodge dual of a KYT $f$ is a closed conformal Killing-Yano tensor\footnote{So a CCKYT 
$h$ satisfies \re{idna} with the first term on the right hand side vanishing identically.} (CCKYT) 
$h$ and vice versa \cite{Cariglia:2003kf}, i.e.
\ber
f = \star h \,, \qquad dh=0 \,.
\eer{ccky}
Specifically when the rank of the CCKYT $h$ is \( n = D-1 \), one has  
\ber
\na_{a} h_{b_{1} \dots b_{D-1}} = (D-1) g_{a[b_{1}} H_{b_{2} \dots b_{D-1}]} \,,
\eer{cckyspe} 
so that e.g. the Hodge dual of the rank-1 current \re{C1} in $D=4$ naturally gives rise to 
another conserved current of rank-2, as mentioned in the context of \re{Pen}. Likewise the 
Hodge dualisation of the rank-1 Cotton currents \re{C2} give rise to two more conserved 
currents of rank-2 and rank-$(D-2)$, respectively. Explicitly, the rank-$(D-2)$ dual of a rank-2 
CKYT $k$ brings about the current
\ber
\mathscr{J}^{a} = C^{abc} \epsilon_{b c d_{1} \dots d_{D-2} } k^{d_{1} \dots d_{D-2}} \,.
\eer{nCotCur}
In three dimensions this relates the Cotton current $\tilde{J}^a$ for a rank-2 CKYT to a 
current for its rank-1 dual defined with the dual of the Cotton tensor, the York tensor.

In a slight detour from the main thrust of this section, we also note that given a CKYT $k$ \re{idna}, 
one may also construct a current from the associated trace $K$ \re{idnb}. For a rank-2 CKYT $k$,
the associated rank-1 trace $K^{a}$ is trivially conserved \( \na_{a} K^{a} = 0 \) \re{idnn}. Moreover,
the ``Einstein current''
\beq 
{\cal J}^{a} := G^{ac} K_{c} \,, \label{EinK}
\eeq 
is also covariantly conserved (with a manifest potential, cf. \re{obstruce}), thanks to 
\begin{equation}
R^{ac} \na_{a} K_{c} = R^{ac} \na_{(a} K_{c)} = 0 \,, \label{symRK}
\end{equation}
derived in \cite{Lindstrom:2021dpm}. When $K^a$ is itself a Killing vector\footnote{As is the 
case for the trace part of a CKYT in $D=4$.}, we again find a dual $(D-1)$-form CCKYT and 
an associated rank-1 current.

The  duality relation \re{ccky} also makes it possible to find a version of the KT-current involving 
CCKYTs. The KT-current \cite{Kastor:2004jk} is defined for a rank-$n$ KYT $f$ as
 \ber
j^{a_{1} \dots a_{n}} = N_{n} \, \delta^{a_{1} \dots a_{n} d_{1} d_{2}}_{b_{1} \dots b_{n} c_{1} c_{2}}
\, f^{b_{1} \dots b_{n}} \, R_{d_{1}d_{2}}{}^{c_{1}c_{2}} \,, 
\eer{KTcom}
where
\( \delta^{a_{1} \dots a_{m}}_{b_{1} \dots b_{m}} = 
\delta^{[a_{1}}_{b_{1}} \cdots \delta^{a_{m}]}_{b_{m}} \)
is the generalised Kronecker delta, which is totally antisymmetric in all up and down indices,
\( N_{n} := - (n+1)(n+2)/4n \) and $R$ is the Riemann curvature tensor. The covariant 
divergence of the KT-current vanishes thanks to the Bianchi identities for $R$. Thus,
dualising $f$ to $h$ via \re{ccky} in \re{KTcom}, we find a new covariantly conserved current 
based on $h$
\ber
j^{a_{1} \dots a_{n}} \sim 
\epsilon^{a_{1} \dots a_{n} b_{1} b_{2}} \, R_{b_{1} b_{2}}{}^{c_{1} c_{2}} \, h_{c_{1} c_{2}}
\sim \star R^{a_{1} \dots a_{n}, c_{1} c_{2}} \, h_{c_{1} c_{2}} \,,
\eer{dudu}
where star denotes the left dual.

The preceding construction involving CCKYTs in the KT-current prompts the question of 
whether one can replace the KYTs in the KT-current by CKYTs, modulo some modifications. 
This turns out not to give a conserved current, as we discuss in the next section.

\section{Rank-\texorpdfstring{$n$}{n} currents from rank-\texorpdfstring{$n$}{n} CKYTs}\label{ConCurn}
In this section, we use some of the formulae derived in section \!\ref{rankn} to investigate a 
rank-$n$ current constructed from a rank-$n$ CKYT. More specifically, we look for a covariantly 
conserved rank-$n$ tensor $J$ linear in the CKYT and its contraction, formed from geometric 
tensors and covariant derivatives, such that $J$ is second-order in covariant derivatives. This 
specification is in line with previous currents constructed using KYTs \cite{Kastor:2004jk, Lindstrom:2021qrk}.

It must be pointed out that there are many options for the terms in such a current and our 
final result, that a large class of them lead to trivial currents, is by no means obvious. It is 
only with the help of a number of identities, some of which are derived in section \!\ref{rankn} 
that we are able to show this (somewhat disappointing) result.

We start by listing a number of relations between terms of the right kind:
\ber
&& R^{[a_{1} | cd | a_{2}} k^{a_{3} \dots a_{n}]}{}_{cd}
= - \half R_{cd}{}^{[a_{1} a_{2}} k^{a_{3} \dots a_{n}] cd} \,, \\[1mm]
&& \na_{c} \na^{[a_{1}} k^{a_{2} \dots a_{n}]c} 
= \na_{c} \na^{[a_{1}} k^{a_{2} \dots a_{n} c]} + (-1)^{n+1} \na^{[a_{1}} K^{a_{2} \dots a_{n}]} \,, 
\label{eq33} \\[1mm]
&& \na^{[a_{1}} \na_{c} k^{a_{2} \dots a_{n}]c}
= (-1)^{n+1} (D+1-n) \na^{[a_{1}} K^{a_{2} \dots a_{n}]} \,, \label{eq34}  \\[1mm]
&& [ \na_{c}, \na^{[a_{1}} ] k^{a_{2} \dots a_{n}]c} = 
(-1)^{n} \frac{(n-1)}{2}R_{cd}{}^{[a_{1} a_{2}} k^{a_{3} \dots a_{n}] cd}
+ R_{c}{}^{[a_{1}} \, k^{a_{2} \dots a_{n}]c} \,,  \label{eq35}  \\[1mm]
&& \Box k^{a_{1} \dots a_{n}} = \na_{c} \na^{[{c}} k^{a_{1} \dots a_{n}]} 
+ n \na^{[a_{1}} K^{a_{2} \dots a_{n}]} \,.
\eer{3dep}
Notice that the difference of \re{eq33} and \re{eq34} gives \re{eq35}, and thus
\ber
F^{a_{1} \dots a_{n}} + (D-n) \na^{[a_{1}} K^{a_{2} \dots a_{n}]}
 = \frac{(n-1)}{2}R_{cd}{}^{[a_{1} a_{2}} k^{a_{3} \dots a_{n}] cd}
+  (-1)^{n} R_{c}{}^{[a_{1}} \, k^{a_{2} \dots a_{n}]c} \,.
\eer{lindep}
These relations let us choose the following rank-$n$, skew-symmetric, independent combinations 
\ber
R_{cd}{}^{[a_{1} a_{2}} \, k^{a_{3} \dots a_{n}]cd} \,, \;
R_{c}{}^{[a_{1}} \, k^{a_{2} \dots a_{n}]c} \,, \; 
R \, k^{a_{1} \dots a_{n}} \,, \;
\na^{[a_{1}} K^{a_{2} \dots a_{n}]}  \,, \;
\eer{Threelist}
as the building blocks of a possible current. For computational ease, without loss of generality, 
we replace the first two terms in the list \re{Threelist} by 
\beq
\breve{{\cal K}}^{a_{1} \dots a_{n}} := 
- \frac{(n-1)}{4} \, R^{[a_{1} a_{2}}\,_{b c} \, k^{a_{3} \dots a_{n}] b c}
+ \frac{1}{2 n} \, R \, k^{a_{1} \dots a_{n}} \,, 
\label{Newcur2}
\eeq
and
\begin{equation}
\hat{{\cal K}}^{a_{1} \dots a_{n}} := R_{c}\,^{[a_{1}} \, k^{a_{2} \dots a_{n}] c} + 
\frac{(-1)^{n}}{n} \, R \, k^{a_{1} \dots a_{n}} \,, 
\label{Newcur1}
\end{equation}
modeled after two conserved currents in \cite{Lindstrom:2021qrk}.

With these preambles, we make an ansatz for the conserved current:
\ber
J^{a_{1} \dots a_{n}} =
2 \breve{{\cal K}}^{a_{1} \dots a_{n}} + \al \hat{{\cal K}}^{a_{1} \dots a_{n}} 
+ \beta R k^{a_{1} \dots a_{n}} + \gamma \na^{[a_{1}} K^{a_{2} \dots a_{n}]} \,.
\eer{link}
To this, we may add any amount of the conserved current $F^{a_{1} \dots a_{n}}$ in \re{theCurn}, 
its coefficient will remain arbitrary.

To proceed, we rewrite the identity \re{BoxLSn} or \re{lindep} in terms of the tensors 
\re{Newcur2} and \re{Newcur1}:
\ber
&& F^{a_{1} \dots a_{n}}
= (-1)^{n} \hat{{\cal K}}^{a_{1} \dots a_{n}} - 2 \breve{{\cal K}}^{a_{1} \dots a_{n}}
+ ( n-D) \na^{[a_{1}} K^{a_{2} \dots a_{n}]} \,.
\eer{Fid}
Using this identity in \re{link} yields for the divergence
\ber\nn
&&
\na_{a_{1}} J^{a_{1} \dots a_{n}}
= (\al + {(-1)^{n}} ) \na_{a_{1}} \hat{{\cal K}}^{a_{1} \dots a_{n}} 
+ (\gamma +n-D ) \na_{a_{1}} \na^{[a_{1}} K^{{a_{2} \dots a_{n}}]}
+ \beta \na_{a_{1}} (R k^{a_{1} \dots a_{n}}) \\[1mm]
&& \qquad \qquad\qquad 
+ \na_{a_{1}} F^{a_{1} \dots a_{n}} \,.
\eer{div3} 
The last term is zero, of course, and the coefficients of the others give \( \al = { (-1)^{n+1}}\), 
\(\beta = 0 \) and \( \gamma = D-n \). Plugging these values into \re{link} and using \re{Fid}, 
the rank-$n$ current is thus shown to be trivial, i.e. proportional to the current 
$F^{a_{1} \dots a_{n}}$ in \re{theCurn}.

We thus see that we cannot construct a nontrivial rank-$n$ conserved current from the ansatz 
\re{link}. This ansatz is quite general and covers, e.g., possible generalisations of the KT-current, 
that is local, geometric, linear in $k$ and second-order in $\na$s.

As an additional illustration of the problem, we further discuss the cases $n=1$ and $n=2$ in 
appendix \ref{specase}.

\section{Conserved charges}\label{charges}
One obvious reason for constructing currents is to use them to find conserved charges. In this 
section we illustrate the procedure for doing that on two well-known solutions. For the case of
the Kerr-Newman metric used in the Einstein current \re{EinK}, we are able to derive a 
closed-form charge expression which reduces to a complicated integral on which we comment.
Unfortunately though, the corresponding expression for the C-metric with the trivial 
current \re{theCurn} diverges.

\subsection{The Kerr-Newman metric} \label{KN}
The current \re{EinK} is an example of a non-trivial current and 
it can be used for defining a charge in the usual way\footnote{Recall what is meant by 
``conservation'' of charge in the classical field theory sense. For example, consider Maxwell 
theory in flat Minkowski space. One defines charge \( Q:= \int d^3x \, J^0 \), so that at $t=x^0=$ 
const. surfaces one has ``conservation''  \( dQ/dt=0 \). So, it is only natural that the 
charge-integral hypersurface is chosen as spacelike, with an everywhere timelike normal.},
e.g. as described in Sec. 3 of \cite{Lindstrom:2021dpm}:
\ber
{\cal Q} := \int_{\Sigma_{t}} d^3 x \, \sqrt{\gamma} \, n_{a} \, {\cal J}^{a} \,.
\eer{Qdef}
Let us see how things go on the example of the Kerr-Newman metric
\ber
ds^{2} = - \frac{Q(r)}{\rho^{2}} \left( dt - a \sin^{2}\theta \, d\varphi \right)^{2} 
+ \frac{\rho^{2}}{Q(r)} dr^{2} + \rho^{2} d\theta^{2} + \frac{\sin^{2}{\theta}}{\rho^2}
\left( a dt - (r^2+a^2) \, d\varphi \right)^2 \,,
\eer{KNmet}
where 
\begin{align}
Q(r) & = \left(r-r_{+}\right) \left(r-r_{-}\right) \,, \quad \mbox{with} \quad 
r_{\pm} := m \pm \sqrt{ m^{2} - a^{2}  - e^{2} - g^{2}} \,, \\
\rho^{2} & = r^{2}+ a^{2} \cos^{2} \theta \,,
\end{align}
with the rank-2 CKYT \cite{Floyd}
\ber 
k = r \, dr \wedge \left[ dt - a \sin^{2}{\theta} d\varphi \right] + 
a \cos{\theta} \sin{\theta} d\theta \wedge 
\left[a \, dt - \left( r^{2} + a^{2} \right) d \varphi \right] \,, 
\eer{KNk2}
that remarkably has \( K^{a} = - (\pa_{t})^{a} \), the timelike Killing vector of \re{KNmet} at the 
same time. These give
\ber
{\cal Q} = 2 \pi \left(e^2+g^2\right) \int_{r_{+}}^{\infty} dr \int_{0}^{\pi} 
\frac{ d\theta \, \sin{\theta} \, \left(a^2 + 2 (r - r_{-}) (r - r_{+}) - a^2 \cos{2 \theta}\right)}
{\left( a^2 + 2 r^2 + a^2 \cos{2 \theta} \right)\left((r - r_{-}) (r - r_{+}) - a^2 \sin^2{\theta} \right)} \,.
\eer{KNchar}
This integral is equivalent to the integral found in Sec. 5.3 of \cite{Lindstrom:2021dpm}, 
but here in a different context obviously. It is convergent and finite, but as discussed in 
\cite{Lindstrom:2021dpm}, even though the $\theta$ integral can be taken exactly, we were 
unable to evaluate the $r$-integral. See Sec. 5.3 of \cite{Lindstrom:2021dpm} for details.

Finally, going back to the trivial rank-2 current \re{theCurn} for the Kerr-Newman metric, 
we immediately see that the CKYT \re{KNk2} is a CCKYT, i.e. \( \na_{[a} k_{bc]} = 0 \) 
identically. So it is not suitable for defining a conserved charge using the trivial current 
\re{theCurn} for $n=2$.

\subsection{The C-metric} \label{Cmet}
The C-metric in ``spherical-type coordinates" reads \cite{Griffiths:2006tk} 
\ber
ds^2 = \Omega^{2}(r,\theta) \left( -Q(r) \, dt^2 + \frac{dr^2}{Q(r)} + \frac{r^2 \, d\theta^2}{P(\theta)} 
+ P(\theta) \, r^2 \sin^{2}{\theta} \, d\phi^2 \right) \,,
\eer{cmet}
where the conformal factor and the metric functions are
\ber
\Omega(r,\theta) = \frac{1}{1+ \alpha r \cos{\theta}} \,, \quad
Q(r) = \left( 1-\frac{2m}{r} \right) (1- \alpha^2 r^2) \,, \quad 
P(\theta) = 1+ 2 \alpha m \cos{\theta} \,.
\eer{cmetfun}
The analytic extension of the C-metric is thought to represent a pair of black holes that accelerate
from each other due to the presence of a string that is represented by a conical singularity 
\cite{Kinnersley:1970zw}. The C-metric \re{cmet} reduces to the Schwarzschild black hole when 
$\alpha=0$ and has an obvious curvature singularity at $r=0$. In fact, the parameter $m>0$ has 
to do with the mass of the source, whereas the parameter $\alpha$, where \( 0 < 2 \alpha m < 1 \), 
can be interpreted as the acceleration of the black hole with an acceleration horizon at 
\( r = 1/\alpha > 2 m \) \cite{Griffiths:2006tk}. 

In what follows, we will take the $t$-coordinate as temporal and the $r$-coordinate as spatial,
for which one needs \( r \in (2m,1/\alpha) \). To stay away from the coordinate poles, we also 
take \( \theta \in (0,\pi) \). Finally, we restrict \( \phi \in (-C \pi, C \pi) \), where the parameter $C$
determines the balance between the deficit/excess angles on the two halves of the symmetry
axis of the C-metric (with $t, r$ kept constant) \cite{Griffiths:2006tk}. Note also that the choice
\beq 
C = \frac{1}{1+ 2 \alpha m} \label{Cchoi}
\eeq
removes the conical singularity at \( \theta = 0 \) \cite{Griffiths:2006tk}. 

The C-metric \re{cmet} admits two rank-2 CKYTs
\begin{align}
k_1 & = \Omega^{3}(r,\theta) \, r^3 \, \sin{\theta} \, d\theta \wedge d\phi \quad \mbox{with} \quad
K_{1} = \alpha \, \partial_{\phi} \,, \label{CKYT1} \\
k_2 & = \Omega^{3}(r,\theta) \, r \, dr \wedge dt \quad \mbox{with} \quad
K_{2} = - \partial_{t} \,. \label{CKYT2}
\end{align}

For the trivial current $F^{ac}$ \re{theCurn}, one may, in analogy to the discussion in subsection
\ref{KN} (see also further details given in Sec. 3 of \cite{Lindstrom:2021dpm}), define the charge
\ber
{\cal Q}^{c} := \int_{\Sigma_{t}} d^3 x \, \sqrt{\gamma} \, n_{a} \, F^{a c} \,.
\eer{vecQk}
Adapted to the C-metric \re{cmet}, a $t = $ const. hypersurface $\Sigma_{t}$ has the following
unit normal vector $n^{a}$ and volume element $\sqrt{\gamma}$
\beq
n^{a} = - \frac{1}{\Omega(r,\theta) \sqrt{Q(r)}} \left( \pa_{t} \right)^{a} \,, \quad
n_{a} dx^{a} = \Omega(r,\theta) \sqrt{Q(r)} \, dt \,, \quad
\sqrt{\gamma} = \frac{ \Omega^{3}(r,\theta) \, r^2 \, \sin{\theta}}{\sqrt{Q(r)}} \,. \label{hypdat}
\eeq
We first find that the integrand in \re{vecQk} vanishes identically for the CKYT $k_1$ \re{CKYT1}.
Using the remaining CKYT $k_2$ \re{CKYT2} and identifying the integrand in \re{vecQk} as 
\( {\cal P}^{c} := \sqrt{\gamma} \, n_{a} F^{a c} \) for convenience, we next find that
\begin{align}
{\cal P}^{r} & = \frac{\alpha r^2 \, \Omega^3 \, \sin{\theta}}{2}  
\left( \cos{\theta} \left( 7 \alpha^2 m r+4 \right) 
+ \alpha (2 m + 4 r + 6 m \cos{2 \theta} + \alpha m r \cos{3 \theta}) \right) \,, \nn \\
{\cal P}^{\theta} & = - 2 \alpha r \, \Omega^3 \, \sin^{2}{\theta} \, \left( 1 + 2 m \alpha \cos{\theta} \right) 
\,. \nn
\end{align}
It easily follows that the nontrivial $\theta$ integration of ${\cal P}^{r}$ over the $(0,\pi)$ interval
vanishes and one is just left with ${\cal P}^{\theta}$. We find
\bea
\int_{-C \pi}^{C \pi} d\phi \int_{0}^{\pi} d\theta {\cal P}^{\theta} = - \frac{2 \pi^2 C}{\alpha r^2} \, 
\left( \frac{m \left(2 \alpha ^2 r^2 \left(2 \sqrt{1-\alpha ^2 r^2}-3\right)-4 \sqrt{1-\alpha ^2 r^2}+4\right)+\alpha ^2 r^3}{\left(1-\alpha ^2 r^2\right)^{3/2}} \right) . \label{Ptheint}
\eea
The final step is the $r$ integration. which formally can be done, as
\begin{align}
%\int_{2m}^{1/\alpha} dr 
\int \frac{ m \left(2 \alpha ^2 r^2 \left(2 \sqrt{1-\alpha ^2 r^2}-3\right)-4 \sqrt{1-\alpha ^2 r^2}+4\right)+\alpha ^2 r^3 }{r^2 \left(1-\alpha ^2 r^2\right)^{3/2}} =
\frac{ 2 m \left(\alpha ^2 r^2+2 \sqrt{1-\alpha ^2 r^2}-2\right)+r}{r \sqrt{1-\alpha^2 r^2}} 
%\Big|^{1/\alpha}_{2m} \,. \nn
\end{align}
Unfortunately though, the right hand side clearly diverges as $r \to 1/\alpha$.

Finally, one may wonder what the Cotton charge of the C-metric \re{cmet} is. Since the Ricci tensor
identically vanishes for the C-metric \re{cmet}, so does the Cotton tensor. So the Killing-Yano Cotton
current and, hence, charge is trivially zero.

\section{Discussion}\label{disc}
After introducing a number of useful identities and relations for CKYTs, including their correct 
transformations under conformal transformations, we have discussed the construction of 
conserved ``currents'', i.e. covariantly divergence-free tensors, constructed out of geometric 
tensors and CKYTs. We found a number of currents such as the Einstein current \re{EinK},
the Cotton current \re{C2} and their related expressions involving duals of the (C)KYTs. Based 
on the Einstein current for the Kerr-Newman metric and a rank-2 trivial current for the $C$-metric, 
we illustrated the construction of charges for General Relativity solutions.

Contrary to expectations, however, we found that naive generalisations of rank-$n$ KYT 
currents, such as the KT-current, to rank-$n$ CKYT currents yield trivial currents \re{theCurn}.
To show this requires a number of the identities and relations which we introduced in 
section \!\ref{ConCurn}.

In hindsight, it is worth mentioning the following regarding the charge ${\cal Q}$ \re{KNchar} 
of the Kerr-Newman metric that we found in subsection \ref{KN}: As argued in 
\cite{Lindstrom:2021dpm}, the Cotton current \( \tilde{J}^{a} \) \re{C2} for 
the Kerr-Newman metric (see Sec. 5.3 of \cite{Lindstrom:2021dpm} for details) can be written 
in terms of a 2-potential $\ell$ as \( \tilde{J}^{a} = \na_{a} \ell^{a c} \), which reduces to 
\( \ell^{a c} = 8 \na^{[a} K^{c]} \) in terms of the notation used in the present work. On the other 
hand, since the curvature scalar $R=0$ and $K^{a}$ is identical to the timelike Killing vector 
of the Kerr-Newman metric \re{KNmet}, the Einstein current ${\cal J}^{a}$ \re{EinK} in fact 
reduces to \( \tilde{{\cal J}}^{a} = R^{ac} K_{c} \) in this case. Now it is a well-known fact that 
\( \na_{c} \left( \na^{[c} K^{a]} \right) = - R^{a c} K_{c} \) for a Killing vector, so the Cotton 
current \( \tilde{J}^{a} \) \re{C2} and the current ${\cal J}^{a}$ \re{EinK} or \( \tilde{{\cal J}}^{a} \) are 
indeed proportional to each other for the Kerr-Newman metric. In that sense, it is interesting 
that the Einstein current ${\cal J}^{a}$ \re{EinK}, \( \tilde{{\cal J}}^{a} \) and the Cotton current 
\( \tilde{J}^{a} \) \re{C2} \cite{Lindstrom:2021dpm} all define the same charge (up to some 
proportionality constant) ${\cal Q}$ \re{KNchar} for the Kerr-Newman metric. In fact, the 
3-surface integral of  any one of these on a $t =$ const. hypersurface, say $\Sigma_{t}$, 
should be proportional, up to some possible constants, to the Komar ``mass"\footnote{See 
e.g. section 11.2 of the seminal work \cite{Wald:1984rg}.}, which is an integral over the ``boundary" 
2-surface, $\partial \Sigma_{t}$ of constant $t$ and $r$, spanned by $\Sigma_{t}$. 

In \cite{Lindstrom:2021dpm} the Cotton current was lifted to supergravity in three dimensions. 
An interesting open problem is to find supergravity versions also of the trivial currents \re{theCurn}.

As a final comment we note that the physical meaning of currents and charges constructed 
from KTs, KYTs and CKYTs is not always obvious, unlike those from Killing vectors which describe 
isometries. For this reason the charges based on such tensors are often said to generate 
hidden symmetries and their study is quite rewarding.

\bigskip

\noindent{\bf Acknowledgments}\\
{\"O}.S. would like to thank D.O. Devecio\u{g}lu for help with xAct, and an austere and grim first
referee who forced us to forge this paper into a more resilient form. The research of U.L. was
supported in part by the 2236 Co-Funded Scheme2 (CoCirculation2) of T\"UB{\.I}TAK 
(Project No:120C067)\footnote{\tiny However 
the entire responsibility for the publication is ours. The financial support received from 
T\"UB{\.I}TAK does not mean that the content of the publication is approved in a scientific 
sense by T\"UB{\.I}TAK.}.  In addition, support from the Leverhulme trust is gratefully acknowledged by U.L.

\bigskip

\appendix
\section{Some special cases of trivial currents}\label{specase}
Since we are considering rank-$n$ forms, we are restricted by the dimension $D$ of the 
underlying manifold. Interesting cases are therefore \( n = 1, \dots, D/2 \) in even and 
\( n=1, \dots, (D-1)/2 \) in odd dimensions (modulo their Hodge duals). Below we comment on 
the special cases $n=1$ and $n=2$. Separately, we also introduce a rank-1 current, construct 
its charge and discuss its relation to the recently constructed Cotton current for the Kerr-Newman
metric.

\subsection{\texorpdfstring{$n=1$}{n=1}}
This is the special case when the CKYT is a CKV. Thus consider the following contravariant vector
\ber
J^{a} = R^{ab} k_{b} + \alpha R k^{a} + \beta \na^{a} K~.
\eer{anJ1}
Then
\ber
\na_{a} J^{a} = k^{a} \na_{a} R \left( \frac{1}{2} + \alpha + \frac{\beta}{2(1-D)} \right) 
+ K R \left( 1 + \alpha D +  \frac{\beta}{(1-D)} \right) \,.
\eer{J1}
Demanding \( \na_{a} J^{a} = 0 \) implies that either ``\( D=2 \,, \beta = 1 + 2 \alpha \) with $\alpha$ 
left arbitrary" or ``\( D \neq 2 \,, \alpha = 0 \,, \beta = D-1 \)". 
However note that when $D=2$, one automatically has \( R_{ab} = (R/2) g_{ab} \).
So one has a conserved vector 
$J^{a}$ for
\ber
& D=2: & J^{a} = R^{ab} k_{b} + \alpha R k^{a} + (1 + 2\alpha) \na^{a} K 
= (1+ 2\alpha) (\na^{a} K + 2 R k^a) \,, \\[1mm]\nn
& D\neq 2: & J^{a} = R^{ab} k_{b} + (D-1) \na^{a} K = - \Box k^{a} + \na^{a} K = - F^{a} \,.
\eer{J1cur}
This follows the pattern described for $D\ne2$ in section \ref{ConCurn}.

Since we are dealing with a CKV, a natural question to ask is on the transformation properties of 
$J^{a}$ under Weyl scaling of the metric \( \tilde{g}_{ab} = \Om^{2} g_{ab} \). A CKV transforms as 
\ber
\tilde{k}^{a} = k^{a} \quad \mbox{and} \quad \tilde{K} = K + \Om^{-1} (\na_{a} \Om) k^{a} \,.
\eer{CKVtra}
The transformations of the Ricci and scalar curvature is given by
\ber
\tilde{R}_{ab} & = & R_{ab} - \Om^{-1} \big( (D-2) \na_{a} \na_{b} \Om + g_{ab} \Box \Om \big) 
\\[1mm]\nn
&& + \Om^{-2} \big( 2 (D-2) (\na_{a} \Om) \na_{b} \Om + (3-D) g_{ab} (\na_{c} \Om) \na^{c} \Om \big) \,,
\\[1mm]\nn
\tilde{R} & = & \Om^{-2} \big( R + 2 (1-D) \Om^{-1} \Box \Om 
+ (D-1) (4-D) \Om^{-2}  (\na_{c} \Om) \na^{c} \Om \big) \,.
\eer{Curvtra}
For $D=2$, these give
\ber
&& \tilde{J}^{a} = \Om^{-2} J^{a} -(1 + 2 \alpha) \Om^{-2} \big( k^{a} \Box \ln{\Om}  
+ \na^{a} ( k^{c} \na_{c} \ln{\Om} ) \big) \,, \\[1mm]\nn
&& \tilde{\na}_{a} \tilde{J}^{a} = - (1+ 2 \alpha) \Om^{-2} \big( 2 K \Box \ln{\Om} 
+ k^{a} \na_{a} \Box \ln{\Om} + \Box (k^{a} \na_{a} \ln{\Om} ) \big) \,.
\eer{Jtra}
So one has \( \tilde{\na}_{a} \tilde{J}^{a} = 0 \) only when \( \alpha = - 1/2 \), for 
which \( J^{a} = 0 \). The $D\neq 2$ case can be obtained from the discussion of the Weyl scalings 
of a generic rank-$n$ current $F$ given in appendix \ref{Weyl}, by setting $n=1$, of course.

\subsection{\texorpdfstring{$n=2$}{n=2}}
For $n=2$, following the general case we find the conserved (trivial) current
\ber
J^{ab} = -\frac{1}{2} \na_{c} \na^{[a}k^{bc]} \,.
\eer{}
As may be seen from the following relations, proven in \cite{Lindstrom:2021dpm} 
\ber
(D-3) G^{ac} K_{c} = \na_{c} \left( G^{b[c} k_{b}{}^{a]} + (D-2) \na^{[a} K^{c]} \right) 
\eer{obstruce}
and 
\ber
\na_{a} \breve{{\cal K}}^{a b} = \na_{a} \hat{{\cal K}}^{a b} + \frac{1}{2} (D-3) G^{b}{}_{c} K^{c} \,,
\eer{Gr}
the case $D=3$ requires special attention. 

When $D=3$, the Riemann tensor can be expressed in terms of the Ricci 
tensor and the curvature scalar only, and one finds \( \breve{{\cal K}}^{a b} = \hat{{\cal K}}^{a b} \) 
identically. This implies 
\ber 
\frac{1}{2} R^{abcd} k_{cd} + R_{c}{}^{[a} k^{b]c} = - \hat{{\cal K}}^{ab} 
= - \na^{[a} K^{b]} + \Box \, k^{ab} \,, 
\eer{wazz}
where the last equality follows from \re{BoxLSn} evaluated for $n=2$. On the other hand, when $D=3$ one has 
\ber 
\na_{c} \left( \hat{{\cal K}}^{ac} + \na^{[a} K^{c]} \right) = 0
\eer{kazz}
from \re{obstruce}. The discrepancy between \re{wazz} and \re{kazz} is accounted 
for by the trivial current \re{theCurn} for $n=2$ also in $D=3$.

\section{Killing-Yano Cotton current}\label{KYCcur}
In \cite{Lindstrom:2021dpm}, we showed that the Killing-Yano Cotton current
\( J^a := C^{abc} k_{bc} \), constructed out of a rank-2 CKYT $k$ and the Cotton 
tensor $C$, is covariantly conserved. It is known that the Cotton tensor transforms as
\ber
\tilde{C}_{abc} = C_{abc} - (D-2) ( \na_{d} \ln{\Om}) W^{d}{}_{abc} 
\eer{Ctf}
under conformal transformations \( \tilde{g}_{ab} = \Om^{2} g_{ab} \). Meanwhile, one 
has\footnote{Here we correct an error in \cite{Lindstrom:2021dpm}, where a factor of 
$(D-1)$ was inadvertently forgotten in the second term on the right-hand-side of 
the second equation in eqn. (2.12) in that paper, with no effect on the ensuing discussion.}
\ber
\tilde{k}_{ab} = \Om^{3} k_{ab}  \quad \mbox{and} \quad 
\tilde{K}_{c} = \Om K_{c} + ( \na_{a} \Om) k^{a}{}_{c} \,.
\eer{k2tf}
These give 
\ber
\tilde{J}^{a} = \Om^{-3} \left( J^{a} - (D-2) ( \na_{d} \ln{\Om}) W^{da}{}_{bc} k^{bc} \right) \,. 
\eer{Jtf}
Using these and the property \( (D-3) C_{bdc} = (D-2) \na_{a} W^{a}{}_{bcd} \),
it is straightforward to show that 
\[ \tilde{\na}_{a} \tilde{J}^{a} = \na_{a} \tilde{J}^{a} + D ( \na_{a} \ln{\Om}) \tilde{J}^{a} = 0 \,, \]
independent of the dimension $D$.

\end{document}